\documentclass[pra,twocolumn,showpacs]{revtex4}
\usepackage{amssymb}
\usepackage{graphicx}

\newcommand{\beq}{\begin{equation}}
\newcommand{\eeq}{\end{equation}}
\newcommand{\beqa}{\begin{eqnarray}}
\newcommand{\eeqa}{\end{eqnarray}}
\newcommand{\ba}{\begin{array}}
\newcommand{\ea}{\end{array}}

\begin{document}

\title{Nearly-one-dimensional self-attractive Bose-Einstein condensates in
optical lattices}
\author{L. Salasnich$^{1,2}$, A. Cetoli$^{2}$, B. A. Malomed$^{3}$ and F.
Toigo$^{1,2}$}
\affiliation{$^1$CNISM, Unit\`a di Padova, Via Marzolo 8, 35131 Padova, Italy \\
$^2$Dipartimento di Fisica ``Galileo Galilei'', Universit\`a di Padova, Via
Marzolo 8, 35131 Padova, Italy \\
$^{3}$Department of Interdisciplinary Studies, School of Electrical
Engineering, Faculty of Engineering, Tel Aviv University, Tel Aviv 69978,
Israel }

\begin{abstract}
Within the framework of the mean-field description, we investigate atomic
Bose-Einstein condensates (BECs), with attraction between atoms, under the
action of strong transverse confinement and periodic (optical-lattice, OL)
axial potential. Using a combination of the variational approximation (VA),
one-dimensional (1D) nonpolynomial Schr\"{o}dinger equation (NPSE), and
direct numerical solutions of the underlying 3D Gross-Pitaevskii equation
(GPE), we show that the ground state of the condensate is a soliton
belonging to the semi-infinite bandgap of the periodic potential. The
soliton may be confined to a single cell of the lattice, or extend to
several cells, depending on the effective self-attraction strength, $g$
(which is proportional to the number of atoms bound in the soliton), and
depth of the potential, $V_{0}$, the increase of $V_{0}$ leading to strong
compression of the soliton. We demonstrate that the OL is an effective tool
to control the soliton's shape. It is found that, due to the 3D character of
the underlying setting, the ground-state soliton collapses at a critical
value of the strength, $g=g_{c}$, which gradually decreases with the
increase of the depth of the periodic potential, $V_{0}$; under typical
experimental conditions, the corresponding maximum number of $^{7}$Li atoms
in the soliton, $N_{\max }$, ranges between $8,000$ and $4,000$. Examples of
stable multi-peaked solitons are also found in the first finite bandgap of
the lattice spectrum. The respective critical value $g_{c}$ again slowly
decreases with the increase of $V_{0}$, corresponding to $N_{\max }\simeq
5,000$.
\end{abstract}

\maketitle


\section{Introduction}

It has been firmly established in experiments with ultracold vapors of
alkali metals that Bose-Einstein condensates (BECs) with weak attractive
interactions between atoms (characterized by negative scattering length) can
form stable matter-wave solitons in nearly one-dimensional (1D)
``cigar-shaped" traps, which tightly confine the condensate in two
transverse directions, but leave it almost free along the longitudinal axis.
This setting has made it possible to create stable bright solitons \cite%
{Strecker02,Khaykovich02} and trains of such solitons \cite{Strecker02} in
the $^{7}$Li condensate, where the interaction between atoms may be made
weakly attractive by means of the Feshbach-resonance technique. In the $^{85}
$Rb condensate trapped under similar conditions, stronger attraction between
atoms leads to the creation of nearly 3D solitons in a post-collapse state
\cite{Cornish}.

This experimentally relevant setting is described by effectively 1D
equations which were derived, by means of various approximations, from the
full 3D Gross-Pitaevskii equation (GPE) \cite{PerezGarcia98}-\cite{Napoli2}.
In some cases, the deviation of the effective equation from the
straightforward one-dimensional GPE with the cubic nonlinearity is
adequately accounted for by an extra self-attractive quintic term, which may
essentially affect the solitons \cite{Shlyap02,Brand06,Lev}. The derivation
of the 1D equation starts with adopting an ansatz factorizing the 3D wave
function into the product of a transverse one (which amounts to the ground
state of the 2D harmonic oscillator) and slowly varying axial
(one-dimensional) wave function. The substitution of this \textit{ansatz} in
the variational approximation (VA; a review of the method can be found in
Ref. \cite{Progress}) leads, without resorting to additional approximations,
to a \textit{nonpolynomial Schr\"{o}dinger equation} (NPSE) for the
longitudinal wave function \cite{sala1,sala2}. This equation was
successfully used in various physical contexts \cite{sala3,sala4}. In
particular, the above-mentioned simplified equation including cubic and
quintic terms can then be obtained by an expansion of the NPSE for the case
of a relatively weak nonlinearity \cite{Lev}. A generalization of the NPSE
for a two-component BEC was recently developed in Ref. \cite{we}, in the
form of a coupled system of NPSEs.

The objective of this work is to analyze the dynamics of quasi-1D solitons
in the cigar-shaped trap equipped with a periodic potential, which can be
easily created as an \textit{optical lattice} (OL), by means of two coherent
counterpropagating laser beams illuminating the condensate (for a recent
review of the BEC dynamics in periodic potentials, see Ref. \cite{Markus}).
Solitons in the 1D nonlinear Schr\"{o}dinger equation (NLS) with the
ordinary self-focusing cubic nonlinearity and periodic (lattice) potential
had been studied some time ago \cite{1Dtheory}. It was shown that a stable
soliton can be trapped by a local potential minimum of the OL. On the other
hand, stable quasi-1D solitons in cigar-shaped traps exist up to a certain
threshold, beyond which they suffer \textit{collapse}. The occurrence of the
collapse in the effective 1D equation reflects the presence of the collapse
in the underlying 3D setting \cite{sala1,sala2,Lev}. A new feature, which
the present paper aims to report and explore, is the influence of the
periodic potential on the collapse threshold for solitons in the quasi-1D
trap. We demonstrate that, for both single-peak solitons found in the
semi-infinite gap of the periodic potential, and multi-peaked solitons found
in finite bandgaps, the collapse threshold gradually \emph{goes down} with
the increase of the potential strength. This feature may be explained by the
fact that a sufficiently strong axial OL potential collects almost all the
condensate in a limited space of a single lattice cell, thus facilitating
the onset of the collapse. The predicted effect should be amenable to
observation by means of standard experimental techniques.

The paper is organized as follows. In Section II, we apply the VA directly
to the description of solitons in the underlying 3D GPE. In this way, the
collapse threshold for the quasi-1D solitons is predicted in a
semi-analytical form [in the same approximation, the stability of the
solitons is estimated by means of the Vakhitov-Kolokolov (VK)\ criterion].
In Section III, we derive the one-dimensional NPSE in the presence of the
periodic potential, and find soliton families as numerical solutions of the
latter equation. Comparison with direct numerical solutions of the
underlying 3D\ GPE demonstrates that the NPSE provides for very high
accuracy in the prediction of both the shape of the solitons belonging to
the semi-infinite gap and their collapse threshold; the accuracy provided by
the VA is lower, but nevertheless reasonable too. In Section IV, we briefly
consider examples of solitons belonging to the first finite bandgap. They
are found in a numerical form from the corresponding NPSE. In the
semi-infinite gap, the solitons may be narrow, occupying, essentially, a
single site of the lattice potential, or broad, extending to several sites,
but they always feature a single tall peak. The soliton in the finite
bandgap has a very different shape, with many peaks. The paper is concluded
by Section V.

\section{Variational approach}

The energy-per-atom ($E$) and chemical potential ($\mu $) of the
self-attractive BEC described by the mean-field stationary wave function, $%
\psi (\mathbf{r})$, in the presence of the strong transverse harmonic
confinement with frequency $\omega _{\bot }$, acting in the plane of $\left(
x,y\right) $, are
\begin{equation}
E=\int d\mathbf{r}\,\psi ^{\ast }(\mathbf{r})\left[ -{\frac{1}{2}}\nabla
^{2}+{\frac{1}{2}}(x^{2}+y^{2})+U(z)-\pi g|\psi (\mathbf{r})|^{2}\right]
\psi (\mathbf{r}),  \label{efun}
\end{equation}%
\begin{equation}
\mu =E-\pi g\int \left\vert \psi (\mathbf{r})\right\vert ^{4}d\mathbf{r.}
\label{mu}
\end{equation}%
Here, the OL potential acting along axis $z$ is%
\begin{equation}
U(z)=-V_{0}\cos {\left( 2k_{L}z\right) ,}  \label{U}
\end{equation}%
with $k_{L}=\left( 2\pi /\lambda \right) \sin \left( \theta /2\right) $,
where $\lambda $ is the wavelength of two laser beams with angle $\theta $
between them that create the OL. We assume normalization $\int |\psi (%
\mathbf{r})|^{2}d\mathbf{r}\equiv 1$, then $g\equiv 2|a_{s}|N/a_{\bot }$ is
the adimensional strength of the self-attraction, with negative scattering
length of atomic collisions $a_{s}$, and the number of atoms in the
condensate, $N$. Lengths in Eqs. (\ref{efun}) - (\ref{U}) are measured in
units of the transverse harmonic length, $a_{\bot }=\sqrt{\hbar /\left(
m\omega _{\bot }\right) }$ ($m$ is the atomic mass), and the depth of the
potential, $V_{0}$, is taken in units of $\hbar \omega _{\bot }$, therefore $%
\omega _{\perp }$ does not appear in the equations. For $^{7}$Li atoms in
the transverse trap with $\omega _{\perp }=2\pi \times 1$~KHz, one has $%
a_{\bot }\simeq $ $1$ $\mu $m; then, $k_{L}=1$ (the value used in exampled
displayed below) and $\lambda =1.5$ $\mu $m correspond to $\theta \simeq
30^{\circ }.$ In the same case, $V_{0}=1$ is tantamount, in physical units,
to $\simeq 1.5E_{\mathrm{rec}}$, where the recoil energy for the OL with
period $d\equiv \lambda /\left( 2\sin (\theta /2\right) )$ is $E_{\mathrm{rec%
}}=\left( \pi \hbar /d\right) ^{2}/m$.

To predict solitons in an approximate analytical form, we use the 3D
Gaussian ansatz,
\begin{equation}
\psi (\mathbf{r})={\frac{1}{\pi ^{3/4}\sigma \eta ^{1/2}}}\exp {\left\{ -{%
\frac{(x^{2}+y^{2})}{2\sigma ^{2}}-\frac{z^{2}}{2\eta ^{2}}}\right\} },
\label{ansatz}
\end{equation}%
where $\sigma $ and $\eta $ are, respectively, the transverse width and
axial length of the localized pattern. Inserting this ansatz into Eqs. (\ref%
{efun}) and (\ref{mu}), we obtain
\begin{equation}
E={\frac{1}{2}}\left( \frac{1}{2\eta ^{2}}+{\frac{1}{\sigma ^{2}}}+\sigma
^{2}\right) -{\frac{g}{2\sqrt{2\pi }}}{\frac{1}{\sigma ^{2}\eta }}-V_{0}\exp
{\left( -k_{L}^{2}\eta ^{2}\right) },  \label{E}
\end{equation}%
\begin{equation}
\mu =E-{\frac{g}{2\sqrt{2\pi }}}{\frac{1}{\sigma ^{2}\eta }.}  \label{muE}
\end{equation}

Aiming to predict the ground state in the framework of the above
approximation, we look for values\ of $\sigma $ and $\eta $ that minimize
energy $E$ [as given by Eq. (\ref{E})], using equations $\partial E/\partial
\sigma =\partial E/\partial \eta =0$. This way, we derive coupled equations,
\begin{equation}
-{\frac{1}{\eta ^{3}}}+{\frac{g}{(2\pi )^{1/2}}}{\frac{1}{\sigma ^{2}\eta
^{2}}}+4V_{0}k_{L}^{2}\eta \exp {\left( -k_{L}^{2}\eta ^{2}\right) }=0,
\label{var1}
\end{equation}%
\begin{equation}
-{\frac{1}{\sigma ^{3}}}+\sigma +{\frac{g}{(2\pi )^{1/2}}}{\frac{1}{\sigma
^{3}\eta }}=0,  \label{var2}
\end{equation}%
which can be solved numerically \cite{method}. Obviously, these solutions
yield a ground state, i.e., a minimum of energy, only if the curvature of
the energy dependence, $E(\eta ,\sigma )$, is positive,
\[
{\frac{\partial ^{2}E}{\partial \eta ^{2}}}{\frac{\partial ^{2}E}{\partial
\sigma ^{2}}}-\left( {\frac{\partial ^{2}E}{\partial \eta \partial \sigma }}%
\right) ^{2}>0\;.
\]%
As concerns the dynamical stability of solitons against small perturbations,
it may be, first of all, estimated by means of the VK\ criterion \cite{VK},
according to which a necessary stability condition is $d\mu /dg<0$ (in the
present notation), if the soliton family is described by dependence $\mu (g)$%
. Actually, the VK criterion may also be sufficient for the stability of
solitons in the GPE with self-attraction and lattice \ potential \cite%
{Salerno}.

In Fig. 1, we plot solutions for $\sigma $ and $\eta $ of Eqs. (\ref{var1})
and (\ref{var2}), together with the corresponding energy-per-particle $E$
[calculated as per Eqs. (\ref{E})], as functions of interaction strength $g$%
, for several fixed values of potential depth $V_{0}$ and wavenumber $%
k_{L}=1 $. Additionally, Fig. 2 displays respective dependences $\mu (g)$ of
the soliton's chemical potential, found from Eq. (\ref{muE}). The figures
show that the soliton in the self-attractive BEC is predicted to exist up to
a critical strength, $g_{c}$, which depends on $V_{0}$ and $k_{L}$. At $%
g>g_{c} $, the 3D collapse of the nearly-1D soliton occurs, as suggested by
results obtained previously in models without the OL potential \cite%
{PerezGarcia98,sala1,sala2}. Note that the VA predicts $g_{c}=1.55$ for $%
V_{0}=0$ , which is somewhat higher than the critical value, $g_{c}=1.33$,
obtained from a numerical solution of the full GPE with $V_{0}=0$ in three
dimensions \cite{gammal,sala2}. The characteristic value of the nonlinearity
strength in Figs. 1 and 2, $g=1$, corresponds, for the above-mentioned
transverse size, $a_{\perp }=1$ $\mu $m, and scattering length $a_{s}=-0.1$
nm, which can be attained in $^{7}$Li by means of the Feshbach resonance
\cite{Strecker02,Khaykovich02}, to solitons built of $N\simeq 5,000$ atoms.

As $g$ drops to zero, the axial size of the soliton, $\eta$, diverges, while
transverse width $\sigma $ approaches $1$ (in physical units, it becomes
equal to the above-mentioned harmonic-oscillator length, $a_{\bot }$). On
the other hand, as $g$ approaches $g_{c}$, both $\eta $ and $\sigma $ remain
finite and smaller than $1$.

\begin{figure}[tbp]
{\includegraphics[height=3.7in,clip]{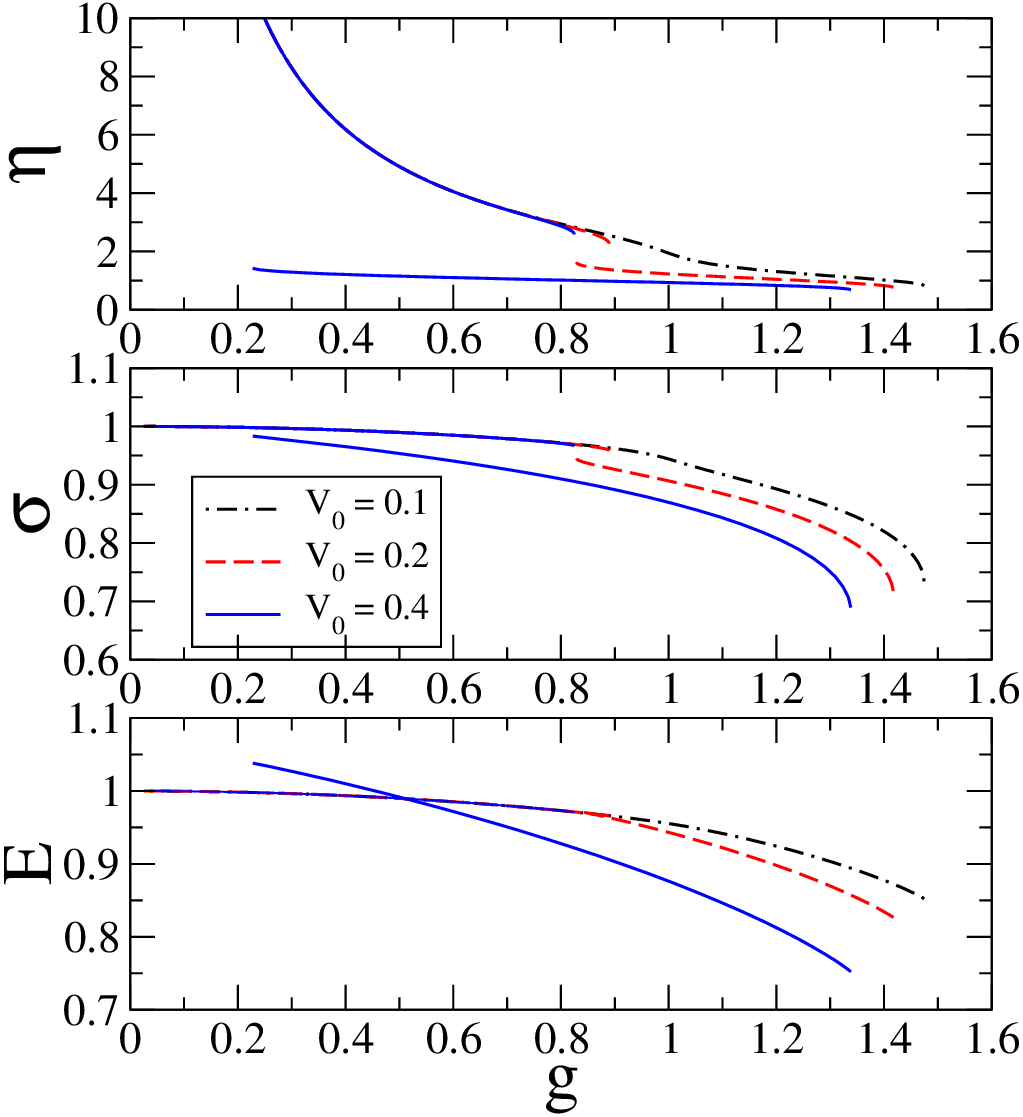}}
\caption{(Color online). Axial length $\protect\eta $, transverse width $%
\protect\sigma $, energy-per-particle $E$ of the quasi-1D soliton versus the
self-attraction strength, $g\equiv 2|a_{s}|N/a_{\bot }$, as predicted by the
variational approximation based on ansatz (\protect\ref{ansatz}) and Eqs. (%
\protect\ref{var1}), (\protect\ref{var2}). The dependences are displayed at
fixed values of depth $V_{0}$ of periodic potential (\protect\ref{U}), with $%
k_{L}=1$. On the right side, all curves terminate at a critical point, $%
g=g_{c}(V_{0})$, beyond which the collapse occurs (the upper curve in the
top panel, which is cut by the panels's frame, actually extends up to $g=0$,
like its counterpart in the bottom panel.}
\label{Fig1}
\end{figure}

\begin{figure}[tbp]
{\includegraphics[height=3.in,clip]{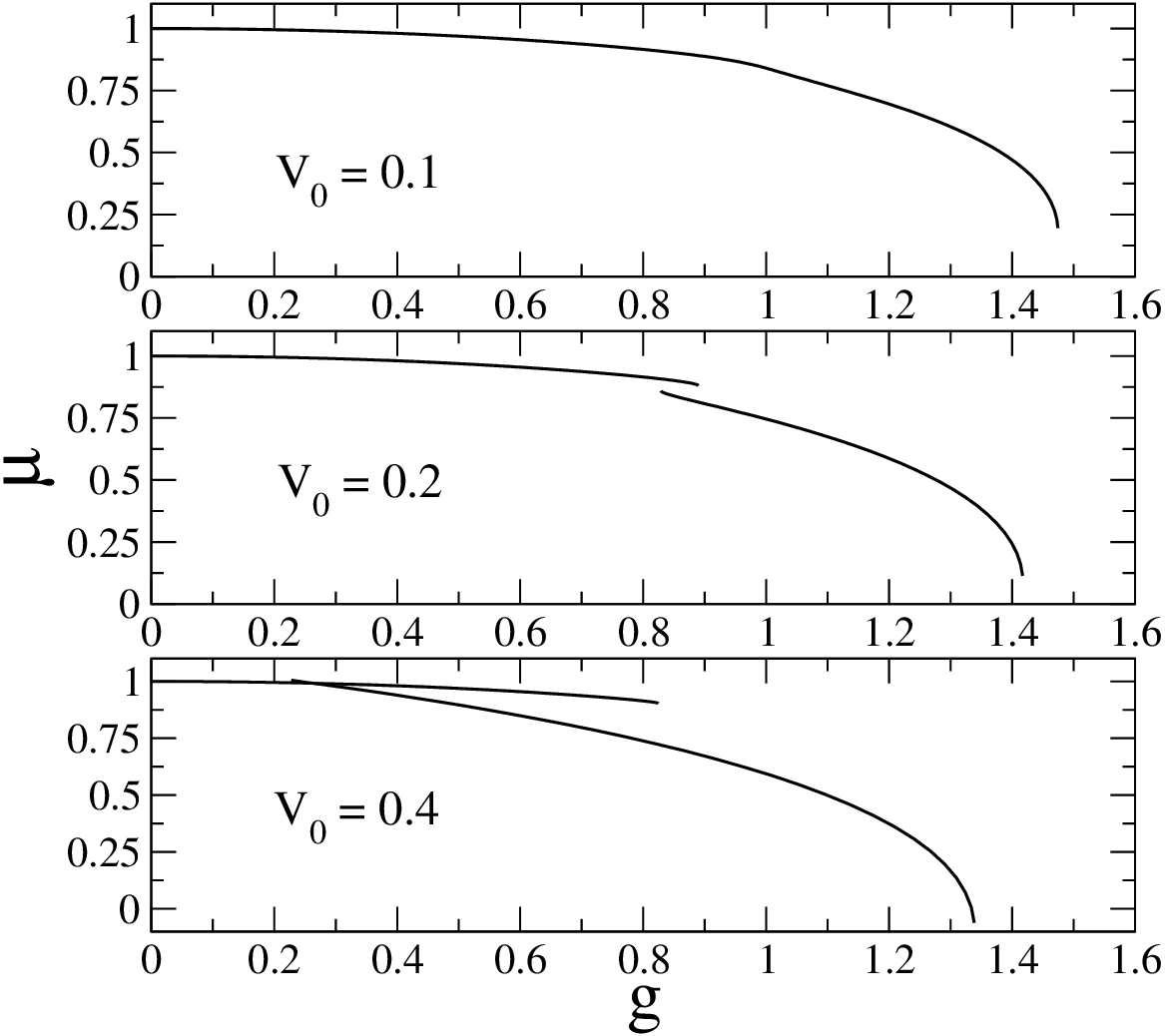}}
\caption{Continuation of Fig. \protect\ref{Fig1}: chemical potential $%
\protect\mu $ of the quasi-1D soliton versus $g$, as predicted by the
variational approximation for the same three values of $V_{0}$, and $k_{L}=1$%
.}
\label{Fig2}
\end{figure}
At small $V_{0}$, the figure shows only a small distortion of the curves,
and a small reduction of $g_{c}$, in comparison with the previously studied
case of $V_{0}=0$. A qualitative change [which is best pronounced in Fig. 2,
in terms of the $\mu (g)$ dependences] is observed at $V_{0}>0.15$ (i.e.,
for $\gtrsim 1,000$ atoms, according to the above estimate), when there
appear \emph{two different} stable branches. For $V=0.2$ and $V_{0}=0.4$,
the lower branches of the $\eta (g)$ and $\sigma (g)$ dependences exists
only in a finite interval, which we denote as
\begin{equation}
g_{m}<g<g_{c},  \label{existence}
\end{equation}%
while the upper branches extend up to $g=0$, i.e., they exist in interval $%
0<g<g_{M}$, with $g_{M}<g_{c}$. Physically, the lower branches (with smaller
values of axial length $\eta $) correspond to an attractive BEC which is
localized, essentially, within a single cell of the OL; accordingly, we call
the corresponding solution a \textit{single-site soliton}. The upper
branches of $\eta (g)$ and $\sigma (g)$ correspond to a weakly localized
solution, which occupies several lattice sites; therefore, we call it a
\textit{multi-site soliton}. The bottom panel of Fig. 1 demonstrates that,
for $V_{0}>0.15$, the multi-site soliton provides for \emph{smaller energy},
i.e., it represents the ground state, at small $g$. At larger $g$, the
single-site soliton becomes the ground state, while its multi-site
counterpart is a metastable state. In particular, the above estimates
demonstrate that, for $V_{0}=0.4$ (which is tantamount to $\simeq 0.6E_{%
\mathrm{rec}}$), the switch from the multi-site ground state to the
single-site one occurs when the number of atoms (in the $^{7}$Li condensate)
attains values $\sim 2,500$.

Figure 2 suggests that the families of soliton solutions are \emph{%
dynamically stable}, as they always meet the VK criterion, $d\mu /dg<0$. In
the case when two solutions exist, i.e., at $V_{0}>0.15$, both of them are
VK-stable, i.e., the ground state and its metastable counterpart alike.

For $V_{0}>0.45$, the numeral solution of Eqs. (\ref{var1}) and (\ref{var2})
demonstrates that the lower threshold $g_{m}$ of existence interval (\ref%
{existence}) for the single-site soliton vanishes, but this is as an
artifact of the VA, which is actually known in other contexts too \cite%
{Salerno}. It is explained by inaccuracy of the Gaussian ansatz in the limit
of weak nonlinearity, i.e., for widely spread small-amplitude solitons. In
this situation, one may, in principle, apply a more sophisticated ansatz,
combining the Gaussian and periodic functions, such as $\cos(2k_Lz)$;
however, the generalized ansatz results in a cumbersome algebra \cite{Arik},
therefore we do not follow this way here.

It is interesting to predict the collapse critical strength, $g_{c}$, as a
function of parameters $V_{0}$ and $k_{L}$ of the axial periodic potential.
In Fig. 3, we display dependence $g_{c}(k_{L})$ predicted by the VA at four
fixed values of $V_{0}$. For given strength $V_{0}$ of the potential, there
exists wavenumber $k_{c}$ at which $g_{c}$ attains its minimum, i.e., the
collapse has the lowest threshold. Further, Fig. 3 shows that this minimum
of $g_{c}$ decreases with the increase of $V_{0}$, which may be understood
as mentioned above: the strong potential tends to squeeze the entire
condensate into a single call of the lattice, which facilitates the onset of
collapse. On the other hand, at large values of $k_{L}$, $g_{c}$ is
asymptotically constant, as the interaction of the condensate with the
short-period OL becomes exponentially weak, see Eq. (\ref{E}), hence it
produces little effect on the collapse threshold.

\begin{figure}[tbp]
{\includegraphics[height=2.3in,clip]{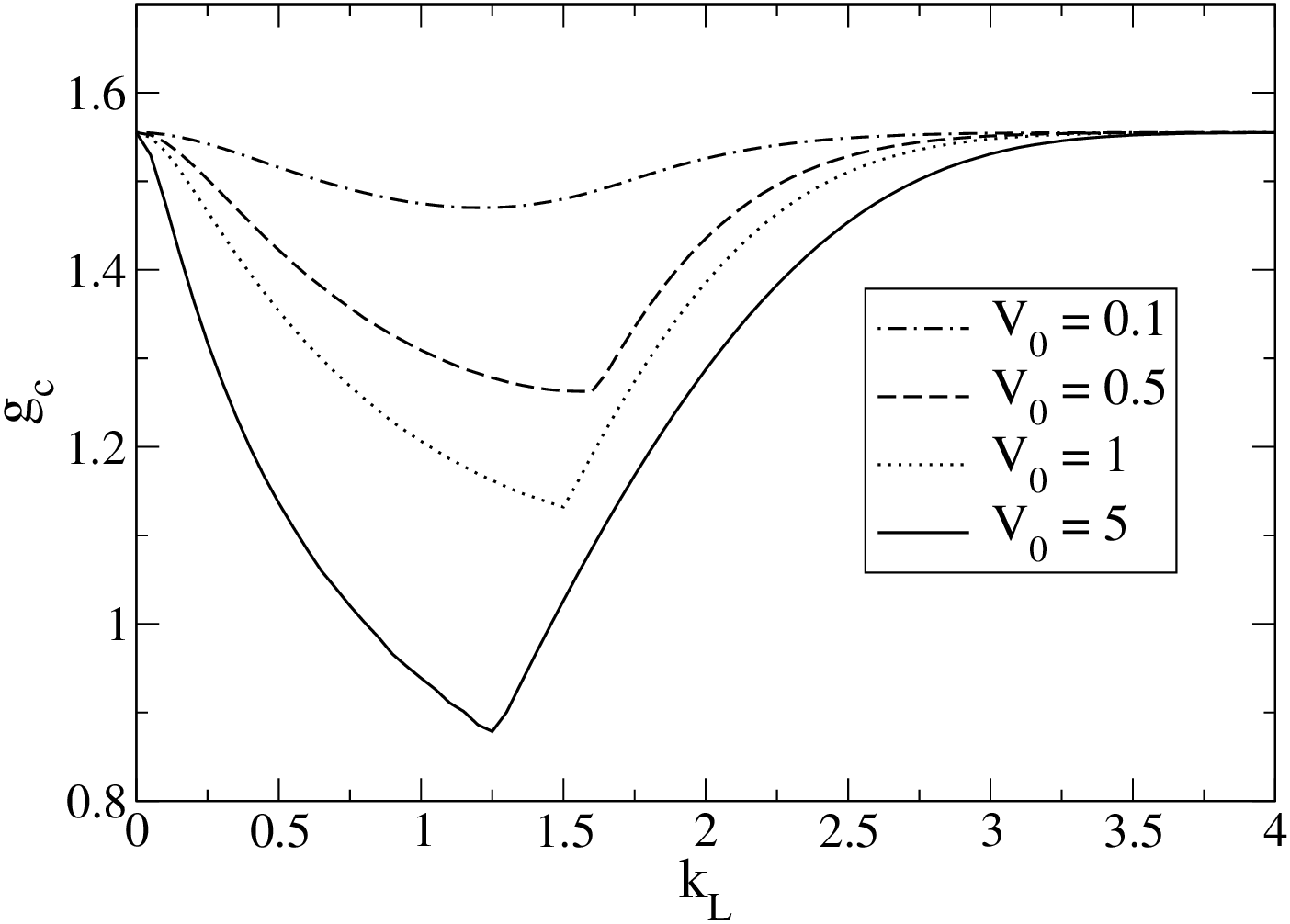}}
\caption{The critical value of the effective self-attraction strength, $%
g_{c} $, above which the soliton collapses, versus wave number $k_{L}$ of
periodic potential (\protect\ref{U}). The results are obtained from the
numerical solution of variational equations (\protect\ref{var1}) and (%
\protect\ref{var2}).}
\label{Fig3}
\end{figure}

\section{Nonpolynomial Schr\"odinger equation}

A more accurate analysis of the present setting may be performed using the
one-dimensional nonpolynomial Schr\"{o}dinger equation (NPSE) \cite{sala1}.
As mentioned in Introduction, the derivation of the NPSE starts with the
ansatz factorizing the 3D wave function into the product of the transverse
2D one, which describes the ground state of the harmonic oscillator, and a
slowly varying 1D axial function, $f(z)$ (which may be complex),
\begin{equation}
\psi (\mathbf{r})={\frac{1}{\pi ^{1/2}\sigma (z)}}\exp {\left\{ -{\frac{%
(x^{2}+y^{2})}{2\sigma (z)^{2}}}\right\} }\,f(z).  \label{ansatz2}
\end{equation}%
Obviously, it is a more general ansatz than the above one, based on Eq. (\ref%
{ansatz}). Inserting expression (\ref{ansatz2}) into Eq. (\ref{efun}), one
finds [upon neglecting spatial derivatives of $\sigma (z)$] the following
one-dimensional energy functional
\[
E=\int dz\,f^{\ast }(z)\left[ -{\frac{1}{2}}{\frac{d^{2}}{dz^{2}}}+{\frac{1}{%
2}}\left( \frac{1}{\sigma (z)^{2}}+\sigma (z)^{2}\right) \right.
\]%
\begin{equation}
\left. +U(z)-{\frac{1}{2}}\frac{g}{\sigma (z)^{2}}|f(z)|^{2}\right] f(z)\;.
\label{e-npse}
\end{equation}%
Further, imposing an extra normalization condition,
\begin{equation}
\int_{-\infty }^{+\infty }dz\,|f(z)|^{2}=1,  \label{N}
\end{equation}%
on the 1D wave function and adding a term with the corresponding Lagrange
multiplier, $\mu $ (which will again be the chemical potential), to energy
functional (\ref{e-npse}), and, finally, minimizing the energy, one arrives
at the following equations for real functions $f(z)$ and $\sigma (z)$ [where
the expression (\ref{U}) for the potential is substituted]:
\begin{equation}
\left[ -{\frac{1}{2}}{\frac{\partial ^{2}}{\partial z^{2}}}-V_{0}\cos {%
(2k_{L}z)}+\frac{1-(3/2)g|f(z)|^{2}}{\sqrt{1-g|f(z)|^{2}}}\right] f(z)=\mu
f(z)\;,  \label{npse}
\end{equation}%
\begin{equation}
\sigma (z)=\left( 1-g|f(z)|^{2}\right) ^{1/4}.  \label{sigma-npse}
\end{equation}%
Equation (\ref{npse}) is the stationary version of the NPSE \cite{sala1}
with the periodic potential, while $\mu $ is fixed by normalization
condition (\ref{N}). Note that only the 1D wave function, $f(z)$, appears in
the NPSE, while the transverse width is locally expressed in terms of $f(z)$%
, as per Eq. (\ref{sigma-npse}). Equation (\ref{npse}) reduces to the
familiar 1D cubic NLS equation for $gf^{2}\ll 1$ [then, Eq. (\ref{sigma-npse}%
) yields $\sigma \approx 1$]. Only in the latter case, the system may be
considered as truly one-dimensional.

In previous studies which used the NPSE \cite{sala1,sala2,sala3,sala4,we},
solitons were not considered in the presence of the periodic potential. In
this work, we solved the full one-dimensional NPSE, which includes the time
derivative and the periodic potential, numerically, using the
finite-difference Crank-Nicholson method in imaginary time \cite%
{sala-numerics}. In this way, we have obtained the soliton profiles shown in
Fig. 4 for different values of the potential depth $V_{0}$ and fixed
self-attraction strength, $g=0.5$ (recall it typically corresponds to $\sim
2,500$ atoms of $^{7}$Li bound in the soliton). As seen from the figure, the
increase of the OL strength from zero to $\simeq 2$ recoil energies may
compress the soliton, in the axial direction, by a factor $\gtrsim 4$.
Generally, the use of the OL offers an efficient means to control the
soliton shape (the OL may also be used as versatile tool to manipulate
matter-wave solitons dynamically \cite{Panos}).

\begin{figure}[tbp]
{\includegraphics[height=2.3in,clip]{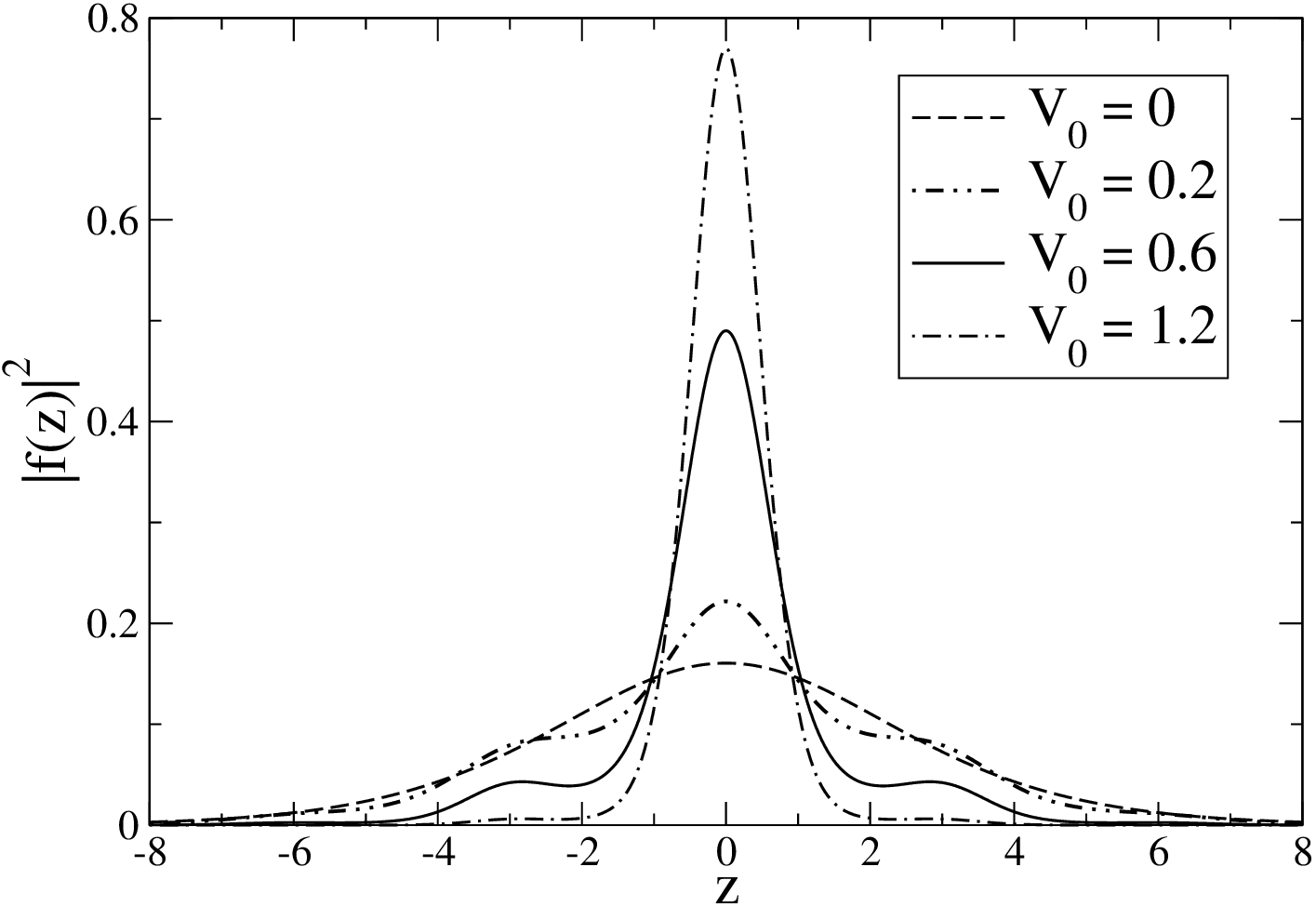}}
\caption{The axial density profile, $|f(z)|^{2}$, of the soliton in periodic
potential (\protect\ref{U}), with $k_{L}=1$ and four different values of $%
V_{0}$. The self-attraction strength is fixed at $g=0.5$.}
\label{Fig4}
\end{figure}
For $V_{0}=0$ (dashed line in Fig. 1), the shape of the soliton has a unique
maximum, and the density profile may be well fitted to $f(z)=\left( \sqrt{{g}%
}{/2}\right) \mathrm{sech}\left( gz{/2}\right) $, which is rigorously valid
in the above-mentioned limit corresponding to $gf^{2}\ll 1$, provided that $%
z $ varies in infinite limits. On the other hand, if $z$ belongs to$\ $a
finite interval, $-L/2<z<L/2$, with periodic boundary conditions, $%
f(z+L)=f(z)$, the NPSE with $V_{0}=0$ yields a spatially uniform profile of
the ground state, $f(z)\equiv 1/\sqrt{L}$, for sufficiently weak
nonlinearity, $0\leq g<\pi ^{2}/L$; the ground state develops a spatial
structure at $g>\pi ^{2}/L$ \cite{ueda,sala4}.

For nonzero but small $V_{0}$ (dot-dot-dashed and solid curves in Fig. 3),
the soliton profile features several local maxima and minima due to the
effect of the periodic potential \cite{notarella}. Thus, under such
conditions, the Bose condensate self-traps into a multi-peaked soliton,
which occupies several cells of the periodic potential (the existence of
\emph{stable three-dimensional} multi-peaked solitons with a similar shape
in the periodic axial potential, but without any transverse confinement, was
demonstrated in Ref. \cite{Warsaw}; in that case, the transverse
self-trapping and stability of the solitons was provided for by the
``Feshbach-resonance-management" technique, i.e., periodic alternation of
the sign of nonlinear coefficient $g$). Following the increase of $V_{0}$,
the soliton compresses in the axial direction, and (for instance, at $%
V_{0}=1.2$, see the dot-dashed curve in Fig. 4) the secondary maxima become
very small. In this case, the condensate actually self-traps into a
single-peak soliton, which occupies only one cell of the OL (\ref{U}).

Contrary to the sharp transition predicted by the VA in the previous
section, the NPSE shows a smooth crossover from the multi-peak soliton to
the single-peak one. In Fig. 5, we plot the axial length, $\sqrt{%
\left\langle z^{2}\right\rangle }$, of the ground-state bright soliton as a
function of strength $g$, for $V_{0}=0.4$ (which is tantamount to the OL
strength $\simeq 0.6$ recoil energy, as shown above), and compare the
prediction of the VA with results produced by the NPSE. It is clearly seen
that, while the VA predicts a jump at $g\simeq 0.5$, the NPSE does not show
it. In fact, this jump is a consequence of the fact that ansatz (\ref{ansatz}%
), which assumes the simple Gaussian waveform for the axial wave function,
provides for a crude fit to multi-peaked states. We also note that the
numerical solution of the NPSE does not reveal the bistability predicted by
the VA, as the numerical algorithm seeks for the ground-state solution for
given $g$. For typical values of physical parameters mentioned above, Fig. 5
implies that, as the number of atoms increases from $1,000$ to $5,000$, the
soliton shrinks in the longitudinal direction from $10$ to $1$ $\mu $m.

\begin{figure}[tbp]
{\includegraphics[height=2.3in,clip]{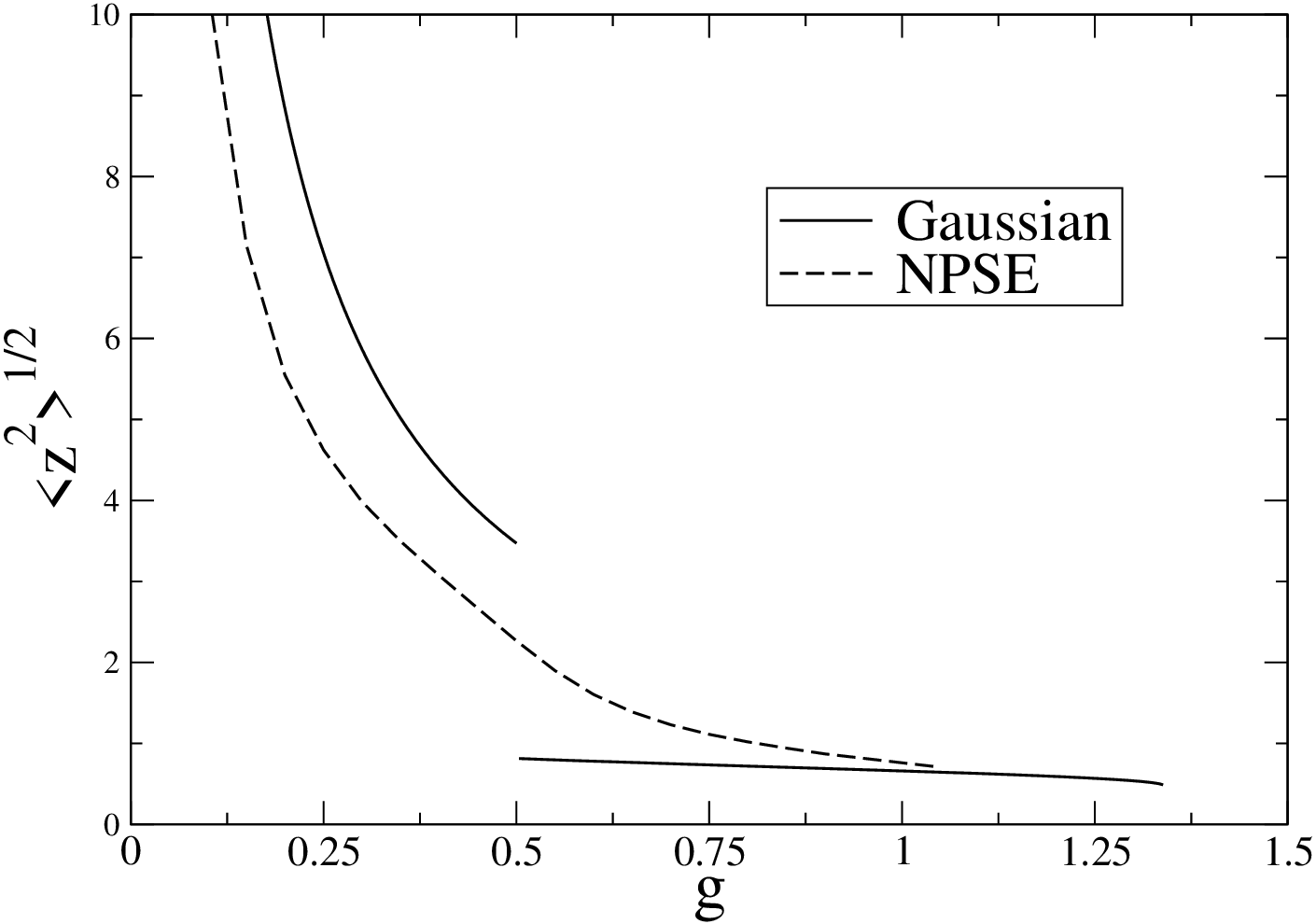}}
\caption{Axial length of the ground-state bright soliton, $\left\langle
z^{2}\right\rangle ^{1/2}$, as a function of self-attraction strength $g$,
for $V_{0}=0.4$ and $k_{L}=1$ in Eq. (\protect\ref{U}). Displayed are
results provided by the variational approximation based on the Gaussian
ansatz, i.e., Eqs. (\protect\ref{var1}) and (\protect\ref{var2}), and by the
nonpolynomial Schr\"{o}dinger equation (NPSE), Eq. (\protect\ref{npse}).}
\label{Fig5}
\end{figure}

The VA developed in the previous section predicts the collapse of the
soliton at the critical value of the self-attraction strength, $g=g_{c}$,
which depends on parameters $V_{0}$ and $k_{L}$ of periodic potential (\ref%
{U}). It is relevant to compare this prediction with results following from
numerical solution of NPSE (\ref{npse}). In Table 1, we present the critical
value, $g_{c}$, found from the NPSE for different values of $V_{0}$ and $%
k_{L}=1$. The results are in qualitative agreement with their variational
counterparts displayed in Fig. 3: the increase of the potential depth, $%
V_{0} $, leads to gradual reduction of $g_{c}$. Note that the value of $%
g_{c}=1.33$ for $V_{0}=0$ is virtually the same as the above-mentioned
numerically exact one, $g_{c}=1.34$, which was found from the numerical
solution of the full three-dimensional GPE (with $V_{0}=0$) \cite%
{sala2,gammal}. According to the above estimates, the drop of $g_{c}$ from $%
1.33$ to $0.85$ corresponds to the decrease of the maximum number of atoms
from $N_{\max }\simeq 8,000$ to $N_{\max }\simeq 4,000$. For the sake of
completeness, the table also shows the axial length and transverse width,
obtained from Eqs. (\ref{npse}) and (\ref{sigma-npse}) at the collapse
point, $g=g_{c}$. It is seen that the soliton shrinks in both directions
with the increase of $V_{0}$, but its length and widths remain finite up to
the collapse point. Typical values of the physical parameters referred to
above imply that the data presented in Table 1 predict roughly constant
density of the condensate at the collapse threshold, $\simeq 3\times 10^{14}$
cm$^{-3}$.

\vskip 0.5cm

\begin{center}
\begin{tabular}{|c|c|c|c|}
\hline\hline
~~~$V_{0}$~~~ & ~~~$g_{c}$~~~~ & ~~~{\small $\sqrt{\left\langle
z^{2}\right\rangle }$}~~~ & ~~~$\sigma(0) $~~~ \\ \hline
0 & 1.33 & 0.91 & 0.75 \\
0.1 & 1.26 & 0.77 & 0.68 \\
0.5 & 1.07 & 0.64 & 0.61 \\
1 & 0.96 & 0.50 & 0.60 \\
2 & 0.85 & 0.41 & 0.57 \\ \hline\hline
\end{tabular}
\end{center}

Table 1. The critical value of the self-attraction strength, $g_{c}$, and
the corresponding values of the axial length, $\sqrt{\left\langle
z^{2}\right\rangle }$, and minimal transverse width, $\sigma (0)$, of the
soliton in the periodic potential, $V(z)=-V_{0}\cos \left( 2k_{L}z\right) $,
with $k_{L}=1$, for different values of $V_{0}$, as found from numerical
solution of the 1D nonpolynomial Schr\"{o}dinger equation, Eq. (\ref{npse}).

\vskip 0.4cm

An important issue is the comparison of the results yielded by the NPSE with
those found from direct numerical solution of the full 3D GPE in the
presence of the axial periodic potential (in previous works dealing with the
NPSE \cite{sala1,sala2,sala3,sala4,we}, such a comparison was not presented
for the model including the OL). It is also interesting to compare the
results with those which can be obtained from the ordinary 1D cubic GPE
[i.e., the 1D equation with the cubic nonlinearity and the same periodic
potential, which can be obtained from Eq. (\ref{npse}) by expanding in small
$g$ and keeping only terms up to the first order in $g$], since the latter
equation is frequently used as a model of the BEC in the quasi-1D traps.
Figure 6 shows that the density profiles generated by the NPSE are always
very close to (practically, coincide with) the ones obtained from the 3D
GPE. On the other hand, the profiles generated by the 1D cubic GPE are
different, and the difference gets more pronounced with the increase of
self-attraction strength $g$. This observation is not surprising because the
nonlinearity in the NPSE essentially deviates from the cubic term if $%
g|f(z)|^{2}$ is large enough. Note also that the cubic GPE in one dimension
cannot predict any collapse.

\begin{figure}[tbp]
{\includegraphics[height=2.3in,clip]{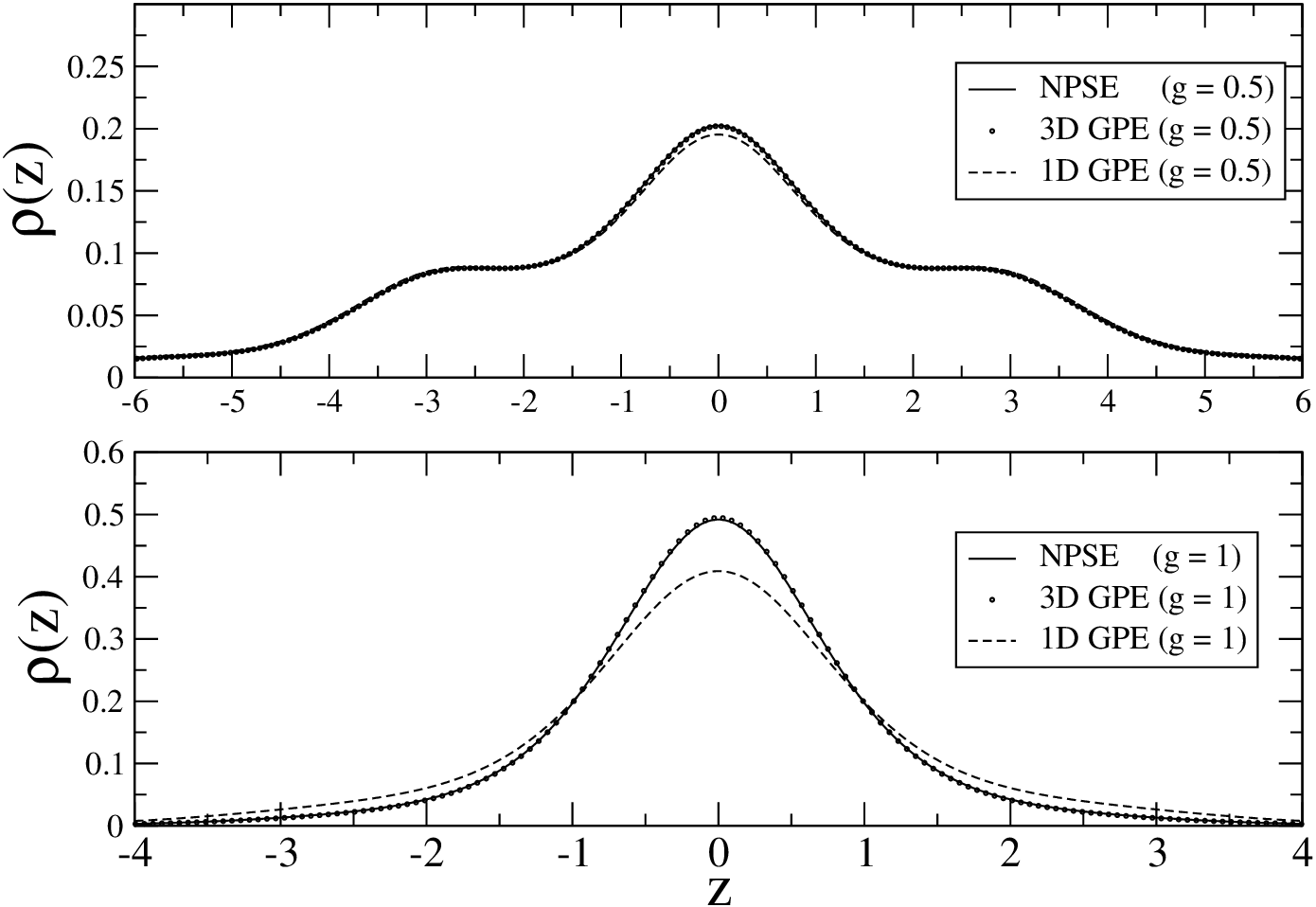}}
\caption{The axial density profile, $\protect\rho (z)$, of the soliton in
potential (\protect\ref{U}), with $k_{L}=1$ and $V_{0}=0.2$. Comparison
between results provided by the different equations: NPSE, 3D GPE, and 1D
GPE (the latter two with the cubic nonlinearity) is presented. In the case
of the 3D equation, Eq. (\protect\ref{efun}), the axial density is defined
as $\protect\rho (z)=\protect\int \protect\int |\protect\psi (\mathbf{r}%
)|^{2}dxdy$, while in the other cases it is simply $|f(z)|^{2}$.}
\label{Fig6}
\end{figure}

\section{Solitons in the finite bandgap}

The above analysis was dealing with solitons whose chemical potential
belongs to the semi-infinite bandgap in the linear spectrum of Eq. (\ref%
{npse}). On the other hand, it is known that the cubic self-attractive
nonlinearity may also give rise to solitons located in higher-order (finite)
bandgaps \cite{Canberra}. This section addresses such solitons in the
present model. It is relevant to stress that up to now, higher-bandgap
solitons, in the case of self-attraction, have been considered only in
strictly one-dimensional settings.

Here, we report soliton solutions found by means of the NPSE, Eq. (\ref{npse}%
), in the first finite bandgap corresponding to $k_{L}=1$. For this purpose,
a self-consistent numerical method was used, with periodic boundary
conditions. We employed a spatial grid of $1025$ points, covering the
interval of $-50.26\leq z\leq 50.26$, which corresponds to $32$ periods of
the external potential. To verify the correctness of the numerical scheme,
we have checked that the lowest-energy state in the semi-infinite bandgap
(i.e., the ground state of the system), produced by this method, is
identical to that found above by the integration of the NPSE in imaginary
time, i.e., the soliton shown in Fig. 4.

As a typical example, in Fig. 7 we plot the density profile, $|f(z)|^{2}$,
of the 33th state (it is number 33 in the full set of the states generated
by the numerical scheme) for $g=0.5$ and three different values of the
potential depth, $V_{0}$. In the linear approximation, this state lies at
the bottom of the second Bloch band. With the increase of nonlinearity
strength $g$, the energy of the 33th state lowers, and, in doing so, it
enters the first finite bandgap from above. At large values of $g$ (and
small values of $V_{0}$), it passes the entire first bandgap from its top to
the bottom, then crosses the band separating this bandgap from the
semi-infinite gap, and eventually sinks into the latter one. Figure 7 shows
that, for $V_{0}=0.2$, this state is still fully delocalized (being similar
to a Bloch wave), while for $V_{0}=0.6$ it becomes localized, with a
mean-square width much smaller than the length of the periodic box,
featuring many local maxima and minima (zeros). We call this solution an
excited multi-site soliton. Its multi-peaked structure is typical to gap
solitons \cite{Canberra,Markus,BBB}. At $V_{0}=1$, Fig. 7 shows that the
excited soliton compresses itself into a narrower state. Simulations of the
NPSE in real-time demonstrate that this soliton family is dynamically stable.

\begin{figure}[tbp]
{\includegraphics[height=2.3in,clip]{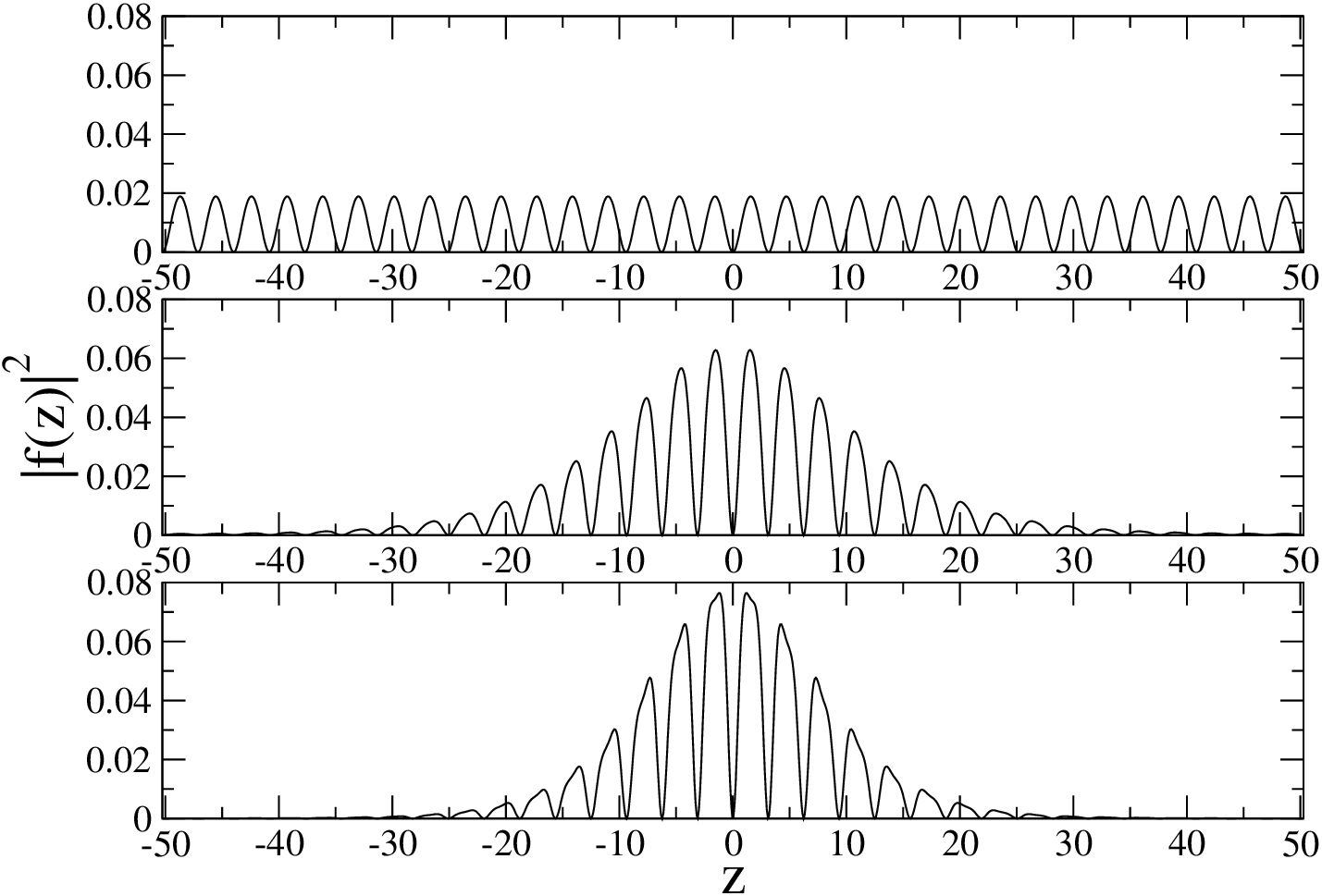}}
\caption{The axial-density profile, $|f(z)|^{2}$, of the lowest state in the
first finite bandgap of periodic potential (\protect\ref{U}) with $k_{L}=1$.
The self-attraction strength is $g=0.5$. The top, middle, and bottom panels
correspond, respectively, to $V_{0}=0.2$, $V_{0}=0.6$, and $V_{0}=1$.}
\label{Fig7}
\end{figure}

\vskip 0.4cm

\begin{center}
\begin{tabular}{|c|c|c|}
\hline\hline
~~~$V_{0}$~~~ & ~~~$g_{c}^{(1)}$~~~~ & ~~~{\small $\sqrt{\left\langle
z^{2}\right\rangle }$}~~~ \\ \hline
4 & 1.16 & 0.70 \\
5 & 1.10 & 0.63 \\
6 & 1.03 & 0.60 \\
7 & 1.00 & 0.57 \\ \hline\hline
\end{tabular}
\end{center}

Table 2. The critical value of the self-attraction strength, $g_{c}^{(1)}$,
of the lowest-energy state in the first finite bandgap of periodic potential
(\ref{U}), with $k_{L}=1$ and increasing values of its depth, $V_{0}$. \ At $%
g>g_{c}^{(1)}$, collapse takes place. The soliton's axial length, $\sqrt{%
\left\langle z^{2}\right\rangle }$, corresponding to to $g=g_{c}^{(1)}$, is
tabulated too. Note that $\sigma (0)$ is always equal to $1$ because the
soliton axial profile $|f(z)|^{2}$ has a node at $z=0$.

\vskip0.4cm

For the lowest-energy state in the first finite bandgap, there also exists a
critical strength, $g_{c}^{(1)}$, of the self-attraction strength, above
which the collapse occurs. In Table 2, we present these critical values
corresponding to the increasing depth of the potential, $V_{0}$, which
demonstrates gradual decrease of $g_{c}^{(1)}$. The latter dependence is
qualitatively the same as reported above for solitons in the semi-infinite
bandgap, cf. Table 1.

The axial size of the soliton at the collapse threshold ($g=g_{c}^{(1)}$) is
also included in Table 1. As for the transverse width, $\sigma (z)$, in all
cases it takes value $\sigma (0)=1$ at the center ($z=0$), unlike the
solitons in the semi-infinite gap, cf. Table 1. This feature is explained by
the fact that the amplitude of the gap soliton vanishes at $z=0$ (see Fig.
7), hence the width of the confined state at this point is the same as in
the linear equation, i.e., $\sigma \equiv 1$ (according to the normalization
adopted above).

In Table 2, we start with $V_{0}=4$ because for smaller $V_{0}$ the
considered state plunges into the semi-infinite bandgap at $g$ close to $%
g_{c}^{(1)}$. According to the above estimates, the largest number of $^{7}$%
Li atoms possible in the soliton created in the first bandgap is $N_{\max
}\sim 5,000$.

For the sake of completeness, in Fig. 8 we present the first $41$
eigenvalues $\epsilon _{j}$, as found from the stationary NPSE equation,
\begin{equation}
\left[ -{\frac{1}{2}}{\frac{\partial ^{2}}{\partial z^{2}}}-V_{0}\cos {(2kz)}%
+{\frac{1-{\frac{3}{2}}g|f_{j}(z)|^{2}}{\sqrt{1-g|f_{j}(z)|^{2}}}}\right]
f_{j}(z)=\epsilon _{j}f_{j}(z)\,,  \label{eigen-npse}
\end{equation}%
and plotted versus interaction strength $g$. The respective eigenfunctions, $%
f_{j}(z)$, may be both delocalized (Bloch-like) and localized
(soliton-like). At $g=0$, the first $32$ eigenvalues generated by our
numerical scheme belong to the first band, while the other nine fall into
the second band. In compliance with the above discussion, Fig. 8 shows that
the lowest eigenvalue and the 33th one split off from the first and second
continuous bands and move down (up to the onset of the collapse) with the
increase of $g$, thus giving rise to localized states, in the semi-infinite
and first finite gaps, respectively. It is noteworthy that the second and
third eigenvalues, which originally belong to the first band, also split off
from it at larger values of $g$ (at largest values of $g$ displayed in Fig.
8, the fourth eigenvalues demonstrates the same trend). We have verified
that the corresponding nonlinear eigenstates become localized, as one may
expect. Note that similar results have been found in Ref. \cite{efremidis}
in a numerical solution of the ordinary cubic GPE in one dimension.

\begin{figure}[tbp]
{\includegraphics[height=2.3in,clip]{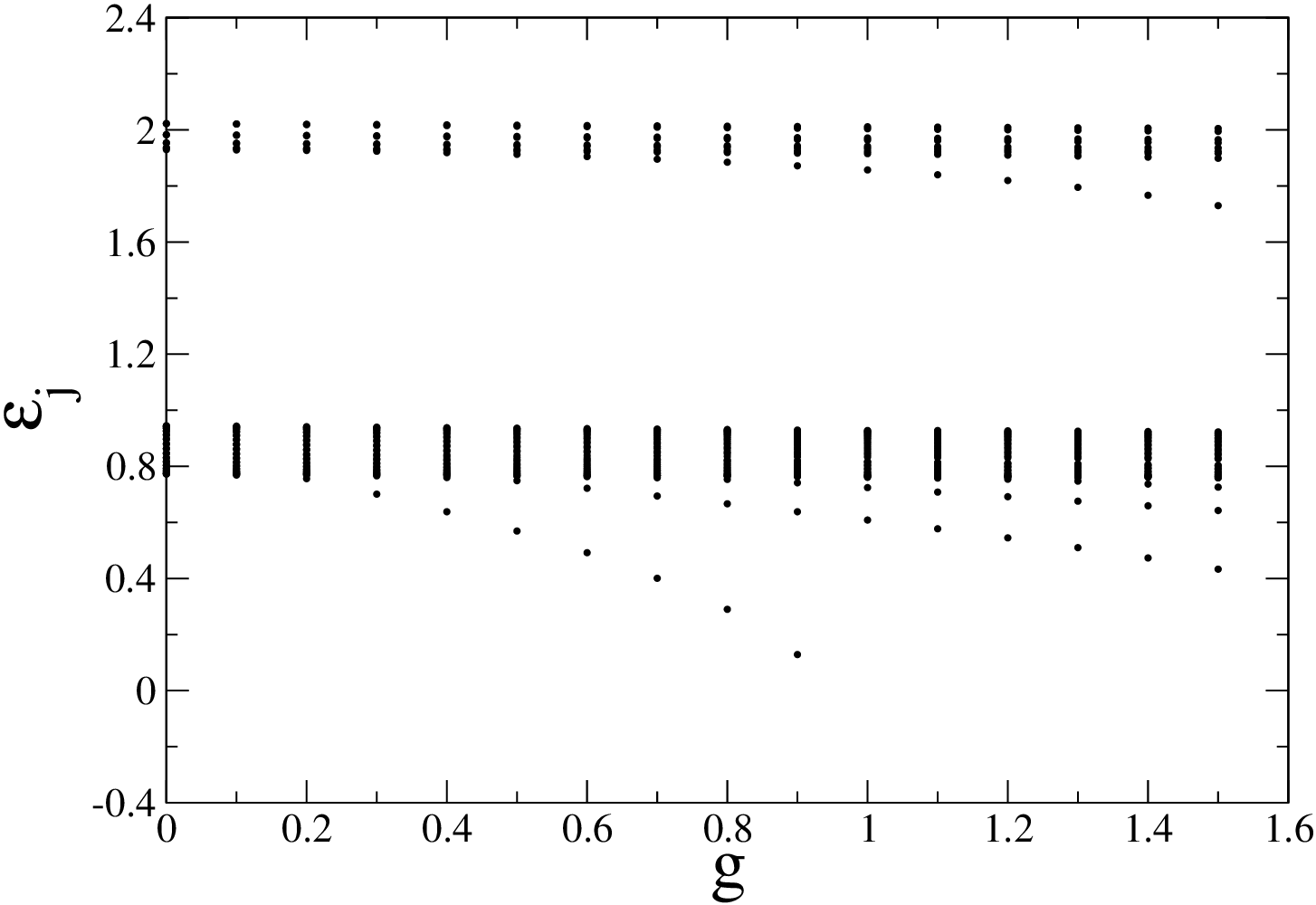}}
\caption{Eigenvalues $\protect\epsilon _{j}$ found from Eq. (\protect\ref%
{eigen-npse}) with $V_{0}=1$ and $k_{L}=1$, as functions of self-attraction
strength $g$.}
\label{Fig8}
\end{figure}

\section{Conclusions}

In this work, we have considered soliton states in self-attractive BECs
loaded into a cigar-shaped trap, which is equipped with a periodic OL
(optical-lattice) axial potential. Using the 3D variational approximation
(VA), and the effective NPSE (nonpolynomial Schr\"{o}dinger equation) in one
dimension, we have demonstrated that the ground state of the condensate is a
bright soliton whose chemical potential falls within the semi-infinite
bandgap created by the periodic potential. With the increase of the
self-attraction strength, $g$, and/or the OL\ strength, $V_{0}$, the
ground-state soliton changes its shape from broad (multi-site) to a narrow
(single-site) one. In particular, the soliton composed of $\sim 2,500$ $^{7}$%
Li atoms trapped in the transverse harmonic potential with $\omega _{\perp
}=2\pi \times 1$~KHz gets compressed by a factor of $4$, as $V_{0}$
increases from zero to two recoil energies; generally, the OL offers a
convenient tool to control the shape of the soliton. The soliton solutions
produced by the VA are stable against small perturbations, according to the
VK\ (Vakhitov-Kolokolov criterion).

Due to the 3D nature of the trapped self-attractive condensate, the solitons
exist up to a critical value, $g_{c}$, of $g$, beyond which the collapse
takes place. A new feature reported in this work is the gradual decrease of $%
g_{c}$ with the increase of $V_{0}$, a qualitative explanation to which was
given. For typical values of experimentally relevant parameters, the results
translate into an estimate for the critical number of atoms at which the
collapse occurs, which ranges from $8,000$ to $4,000$, while the critical
density of the collapsing condensate is, roughly, constant, $\simeq 3\times
10^{14}$ cm$^{-3}$. Comparison with numerical solutions of the underlying 3D
Gross-Pitaevskii equation shows that the NPSE very accurately predicts both
the shape of the ground-state solitons and dependence $g_{c}(V_{0})$; the
accuracy of the variational approximation is somewhat lower, but also
reasonable.

Stable multi-peaked solitons were also found in the first finite bandgap of
the OL potential. The respective critical nonlinearity strength again slowly
decreases with the increase of $V_{0}$, the corresponding largest number of
atoms in the soliton being estimated as $\sim 5,000$. It was studied in
detail how the localized state splits off from the Bloch band and gives rise
to the soliton in the first finite bandgap.

Finally, \ it is relevant to mention that an interesting possibility, which
deserves further consideration, is to induce an \textit{effective} axial
potential not directly, but rather through periodic modulation of the
transverse tight-binding frequency \cite{Napoli}. Another promising
extension of the present analysis may be to the case of \textit{%
self-repulsive} BEC loaded in the combination of the cigar-shaped trap and
axial periodic potential. The study of the latter setting may shed new light
on the theoretical description of quasi-1D gap solitons \cite{Markus,Markus0}%
. These issues will be considered elsewhere.

\section*{Acknowledgement}

The work of B.A.M. was supported, in a part, by the Israel Science
Foundation through the Center-of-Excellence grant No. 8006/03. L.S. thanks
Alberto Parola and Luciano Reatto for useful discussions and suggestions.


\begin{thebibliography}{99}
\bibitem{method} Equations (\ref{var1}) and (\ref{var2}) are handled as a
system of parametric equations, $\sigma =\sqrt{\left( \sqrt{B^{2}+4}%
-B\right) /2}$ and $g=B\sigma \eta /\sqrt{2\pi }$, with parameter $B\equiv
1/\eta ^{2}-4V_{0}k_{L}^{2}\eta ^{2}\exp {\left( -k_{L}^{2}\eta ^{2}\right) }
$.
\end{thebibliography}
\end{document}